\documentclass[12pt,a4paper]{article}
\usepackage{times}
\usepackage{a4wide}
\usepackage{amsfonts}
\usepackage{amssymb}
\usepackage{amsmath}
\usepackage{ifpdf}
\ifpdf
\usepackage[pdftex,unicode,implicit]{hyperref}
\hypersetup{%
  pdftitle    = {A note on the hidden conformal structure of non-extremal black holes},
  pdfkeywords = {gravity, conformal symmetry, quantum gravity, black hole, entropy},
  pdfauthor   = {Tom\'as Ort\'{\i}n and Carlos S. Shahbazi},
  plainpages  = true,
  colorlinks  = true,
  citecolor   = blue,
  urlcolor    = red,
  linkcolor   = black
}
\newcommand{\hepth}[1]{arXiv:{\tt
\href{http://www.arXiv.org/abs/hep-th/#1}{hep-th/#1}}}

\newcommand{\arxiv}[1]{{\tt
\href{http://www.arXiv.org/abs/#1}{arXiv:#1}}}
\else
  \usepackage[dvips]{graphicx}
  \usepackage[unicode,implicit]{hyperref}
  \newcommand{\hepth}[1]{arXiv:{\tt hep-th/#1}}

  \newcommand{\arxiv}[1]{{\tt arXiv:#1}}
\fi
\makeatletter
\@addtoreset{equation}{section}
\makeatother

\begin{document}

\begin{flushright}
\small
IFT-UAM/CSIC-12-36\\
April 20\textsuperscript{th}, 2012\\
\normalsize
\end{flushright}

\begin{center}

\vspace{1cm}

{\LARGE {\bf A note on the hidden conformal structure\\[8mm] of non-extremal black
    holes}}

\vspace{2.5cm}

\begin{center}

\renewcommand{\thefootnote}{\alph{footnote}}
{\sl\large Tom\'{a}s Ort\'{\i}n}
\footnote{E-mail: {\tt Tomas.Ortin [at] csic.es}}\, 
{\sl\large and C. S.~Shahbazi}
\footnote{E-mail: {\tt Carlos.Shabazi [at] uam.es}}
\renewcommand{\thefootnote}{\arabic{footnote}}

\vspace{1.5cm}

{\it Instituto de F\'{\i}sica Te\'orica UAM/CSIC\\
C/ Nicol\'as Cabrera, 13--15,  C.U.~Cantoblanco, 28049 Madrid, Spain}\\

\vspace{4cm}

\end{center}

{\bf Abstract}

\begin{quotation}

  {\small 
    We study, following Bertini \emph{et al.}  \cite{Bertini:2011ga}, the
    hidden conformal symmetry of the massless Klein-Gordon equation in the
    background of the general, charged, spherically symmetric, static
    black-hole solution of a class of $d$-dimensional Lagrangians which
    includes the relevant parts of the bosonic Lagrangian of any ungauged
    supergravity. We find that a hidden $SL\left(2,\mathbb{R}\right)$ symmetry
    appears at the near event- and Cauchy-horizon limits. We extend the two
    $\mathfrak{sl}(2)$ algebras to two full Witt algebras (Virasoro algebras
    with vanishing central charges). We comment on the implications of the
    possible existence of an associated quantum conformal field theory.
}

\end{quotation}

\end{center}

\setcounter{footnote}{0}

\newpage
\pagestyle{plain}



\section*{Introduction}

A complete microscopic explanation of the entropy of an arbitrary black hole
remains as an outstanding challenge for Theoretical Physics. In the mid 90's,
the microscopic degrees of freedom of a charged, static, extremal, black hole
in 5 non-compact dimensions where explicitly identified in the framework of
String Theory, in complete agreement with the Bekenstein-Hawking entropy
\cite{Strominger:1996sh}, providing a first breakthrough in this quest. The
microscopy entropy of many other 4- and 5-dimensional black holes has been
computed successfully following the same pattern.

Although these results were initially thought to depend on the specific
features of String Theory, it has become clear that this is not the case and
the UV details are not important in order to just understand the area law from
a microscopic point of view. The existence of a UV completion, although
important from a fundamental point of view, seems to be irrelevant for this
purpose.  This irrelevance strongly suggests the existence of an universal
underlying principle, which is included in, but not exclusive of String Theory,
which justifies these calculations\footnote{For a recent and comprehensive
  review of these ideas see \cite{Compere:2012jk}.}.

A major step in this direction was taken in \cite{Strominger:1997eq} with the
study of the $(2+1)$-dimensional BTZ black hole \cite{Banados:1992wn}: it had
been already shown in \cite{Brown:1986nw} that the asymptotic symmetry algebra of
this solution was a Virasoro algebra; therefore any consistent quantum theory
describing this black hole should be a conformal field theory, and hence the
Cardy formula can be used to compute the asymptotical growth of states,
obtaining a result that is in agreement with the Bekenstein-Hawking
entropy. This analysis (with some differences, such as considering the
symmetries in the near horizon limit) has been extended to other,
higher-dimensional black holes \cite{Carlip:1998qw}, and a seemingly
universal characteristic of all the black holes whose entropy has been
computed microscopically has emerged: they all are described by 2-dimensional
conformal theories, at least in some limit.

A considerable effort has been dedicated to unveil the hidden conformal
symmetries of the near-horizon region of different kinds of black holes. For
instance, in \cite{Guica:2008mu}, a duality between the extremal Kerr black
hole and a chiral 2-dimensional conformal theory was found. For the
non-extremal Kerr black hole, a different approach has been adopted in
\cite{Castro:2010fd}, where the massless Klein-Gordon equation was used in
order to elucidate the hidden conformal symmetry. In particular, it was shown
that it is possible to define a set of vector fields of a particular
submanifold of the space-time, such as they obey the $\mathfrak{sl}(2)$
algebra and the Casimir gives the massless wave operator. This approach has
later been used in \cite{Bertini:2011ga} and \cite{Wang:2010qv} (see also
\cite{Lowe:2011aa}) for the Schwarzschild and the Kerr-Newman black holes,
respectively\footnote{Previous, closely related results were published in
  \cite{Camblong:2004ye}.}, and it is the one that we are going to use for
general $d$-dimensional black holes in this note, using the metrics introduced
in \cite{Ferrara:1997tw} and \cite{Meessen:2011bd}\footnote{The search for the
  hidden conformal symmetry in static black holes has a long history. See, for
  example, \cite{Govindarajan:2000ag,Birmingham:2001qa,Gupta:2001bg}, and,
  more recently, \cite{Franzin:2011wi}.}.

The note is organized as follows: in Section~\ref{sec:backgroundmetric} we
present the theories that we consider and the generic black-hole metrics that
we will use as a background for the massless Klein-Gordon equation. In
Section~\ref{sec:KG} we will study of the hidden conformal symmetry in the
near-horizon regions (inner and outer) of the 4-dimensional case. The
$d$-dimensional generalization is made in the next Section and we discuss our
results in Section~\ref{sec-discussion}.


\section{The background metric}
\label{sec:backgroundmetric}

We are going to consider black-hole solutions of 4-dimensional theories of the
general form
\begin{equation}
\label{eq:generalaction4}
I
=
\int d^{4}x \sqrt{|g|}
\left\{
R +\mathcal{G}_{ij}(\phi)\partial_{\mu}\phi^{i}\partial^{\mu}\phi^{j}
+2 \Im{\rm m} \mathcal{N}_{\Lambda\Sigma}
F^{\Lambda}{}_{\mu\nu}F^{\Sigma\, \mu\nu}
-2 \Re{\rm e} \mathcal{N}_{\Lambda\Sigma}
F^{\Lambda}{}_{\mu\nu}\star F^{\Sigma\, \mu\nu}
\right\}\, ,   
\end{equation}
which includes the bosonic sectors of all 4-dimensional ungauged
supergravities for appropriate $\sigma$-model metrics $\mathcal{G}_{ij}(\phi)$
and kinetic matrices $\mathcal{N}_{\Lambda\Sigma}(\phi)$ with
negative-definite imaginary part. The indices $i,j,\dotsc$ run over the scalar
fields and the indices $\Lambda,\Sigma,\dotsc$ over the 1-form fields. Their
numbers are related only for $N\geq 2$ supergravity theories.

The metrics of all spherically symmetric, static, black-hole solutions of the
action Eq.~(\ref{eq:generalaction4}) have the general form
\cite{Ferrara:1997tw}
\begin{equation}
\label{eq:generalbhmetric4}
\begin{array}{rcl}
ds^{2} 
& = & 
e^{2U} dt^{2} - e^{-2U} \gamma_{\underline{m}\underline{n}}
dx^{\underline{m}}dx^{\underline{n}}\, ,  \\
& & \\
\gamma_{\underline{m}\underline{n}}
dx^{\underline{m}}dx^{\underline{n}}
& = & 
{\displaystyle \left(\frac{r_{0}}{\sinh{r_{0}\tau}}\right)^{2}}
\left[
{\displaystyle \left(\frac{r_{0}}{\sinh{r_{0}\tau}}\right)^{2}} d\tau^{2} 
+
d\Omega^{2}_{(2)}
\right]\, ,\\
\end{array}
\end{equation}
where $r_{0}$ is the non-extremality parameter and $U(\tau)$ is a function of
the radial coordinate $\tau$ that characterizes each particular solution. In
these coordinates the exterior of the event horizon is covered by the negative
values of $\tau$, the event horizon being located at $\tau\rightarrow -\infty$
and spatial infinity at $\tau \rightarrow 0^{-}$. The interior of the Cauchy
horizon (if any) is covered by part of the positive values of $\tau$, the
inner horizon being located at $\tau\rightarrow +\infty$ while the singularity
is located at some finite, positive, value of $\tau$ 
\cite{Galli:2011fq}.

The last term in the action Eq.~(\ref{eq:generalaction4}) can only occur in
$d=4$ dimensions. Therefore, in the general $d$-dimensional case we shall
consider the Lagrangian
\begin{equation}
\label{eq:generalactiond}
I= \int d^{d}x \sqrt{|g|}\,
\left\{
R  +\mathcal{G}_{ij}(\phi)\partial_{\mu}\phi^{i}\partial^{\mu}\phi^{j} 
+2I_{\Lambda\Sigma}(\phi)F^{\Lambda}{}_{\mu\nu}F^{\Sigma\, \mu\nu}
\right\}\, ,
\end{equation}
where $I_{\Lambda\Sigma}(\phi)$ is an invertible, negative-definite,
scalar-dependent matrix. The metrics of the spherically symmetric, static,
black-hole solutions of (\ref{eq:generalactiond}) have the general form
\cite{Meessen:2011bd}
\begin{equation}
\label{eq:generalbhmetricd}
\begin{array}{rcl}
ds^{2} 
& = &
e^{2U}dt^{2} - e^{-\frac{2}{d-3}U}\gamma_{\underline{m}\,
 \underline{n}}dx^{\underline{m}} dx^{\underline{m}}\, , 
\\
& & \\
\gamma_{\underline{m}\,\underline{n}}dx^{\underline{m}} dx^{\underline{m}}
& = &
{\displaystyle
\left(\frac{\mathcal{B}}{\sinh{\mathcal{B}\rho}} \right)^{\frac{2}{d-3}}  
\left[
\left(\frac{\mathcal{B}}{\sinh{\mathcal{B}\rho}} \right)^{2}
\frac{d\rho^{2}}{(d-3)^{2}}
+d\Omega^{2}_{(d-2)}\right]
}
\, .\\
\end{array}
\end{equation}
Here $\mathcal{B}$ is the the higher-dimensional generalization of the
non-extremality parameter $r_{0}$ and the metric is well defined and covers
the exterior of the event horizon for positive values of $\rho$, the event
horizon being at $\rho \rightarrow +\infty$ and spatial infinity at
$\rho\rightarrow 0^{+}$.

If the above metric describes the exterior of a regular black hole, one can
find from it the metric that covers the interior of the Cauchy horizon (if
any) that metric according to \cite{Meessen:2012su}
\begin{equation}
  \rho \longrightarrow -\varrho\, ,
  \hspace{1cm}
e^{-U^{(+)}(\rho)} \equiv  e^{-U(\rho)} 
\longrightarrow -e^{-U(-\varrho)} \equiv -e^{-U^{(-)}(\varrho)}\, .
\end{equation}
The new metric, determined by the function $U^{(-)}$ has the same general form
in terms of the coordinate $\varrho$ which now takes values in the range
$\varrho\in (\varrho_{\rm sing},+\infty)$ because the metric will generically
hit a singularity before $\varrho$ reaches $0$: if the original $e^{-U^{(+)}}$
is always finite for positive values of $\rho$, the transformed one will have
a zero for some finite positive value of $\varrho$.

In the 4-dimensional case, the area of a 2-sphere at fixed radial coordinate
$\tau=\tau_{0}$ is given by
\begin{equation}
\label{eq:areatau}
A(\tau_{0}) = 4\pi f^2(\tau_{0}) e^{-2 U(\tau_{0})}\, ,
\end{equation}
where
\begin{equation}
\label{eq:ftau}
f(\tau)\equiv \frac{r_{0}}{\sinh{r_{0}\tau}}\, .  
\end{equation}
Therefore, the areas of the event and Cauchy horizons, $A_{+}$ and $A_{-}$,
respectively, are given by
\begin{equation}
\label{eq:areahorizons}
A_{\pm}= \lim_{\tau_{0}\rightarrow \mp \infty} A(\tau_{0}) \, .
\end{equation}

In the $d$-dimensional case, we can write a common expression for the area of
a $(d-2)$-sphere at fixed radial coordinate $\rho=\rho_{0}>0$ in the exterior
of the event horizon or $\varrho=\rho_{0}>0$ in the interior of the Cauchy
horizon:
\begin{equation}
\label{eq:arearho}
A(\rho_{0}) = C_{d-2}  
\left| e^{-U^{(+)}(\rho_{0})} g(\rho_{0})\right| ^{\frac{d-2}{d-3}}\, ,
\end{equation}
where 
\begin{equation}
C_{(d-2)}=\frac{2\pi^{\frac{d-1}{2}}}{\Gamma\left(\frac{d-1}{2}\right)}\, ,
\end{equation}
is the volume of the round $(d-2)$-sphere of unit radius and 
\begin{equation}
\label{eq:grho}
g(\rho)\equiv \frac{\mathcal{B}}{\sinh{\mathcal{B}\rho}}\, .  
\end{equation}
The area of the outer ($+$) and inner ($-$) horizons, $A_{\pm}$ are given by
\begin{equation}
\label{eq:areahorizond}
A_{\pm}= \lim_{\rho_{0}\rightarrow \pm \infty} A(\rho_{0}) \, ,
\end{equation}

We will use Eqs. (\ref{eq:areahorizons}) and (\ref{eq:areahorizond}) later in
order to interpret the near-horizon limits of the massless Klein-Gordon
equation.


\section{The massless Klein-Gordon equation in a general 
static black hole  background}
\label{sec:KG}

In \cite{Bertini:2011ga} it was shown that the massless Klein-Gordon equation
in the background of a 4-dimensional black hole exhibits a $SL(2,\mathbb{R})$
invariance in the near-horizon limit which extends to spatial infinity at
sufficiently low frequencies.  Here we will generalize these results to the
charged, static, spherically symmetric black-hole solutions of 4-dimensional
theories of the form Eq.~(\ref{eq:generalaction4}), with metrics of the
general form Eq.~(\ref{eq:generalbhmetric4}).

In the space-time background given by the metric (\ref{eq:generalbhmetric4}),
the massless Klein-Gordon equation
\begin{equation}
\label{eq:KG}
\frac{1}{\sqrt{|g|}}\partial_{\mu}\left(\sqrt{|g|}g^{\mu\nu}\partial_{\nu} \Phi\right)=0\, ,
\end{equation}
can be written in  the form
\begin{equation}
\label{eq:KG4d}
e^{-2 U}\partial^2_{t}\Phi-e^{2U}f^{-4}\partial^2_{\tau}\Phi 
-e^{2U} f^{-2}\Delta_{S^2}\Phi=0\, ,
\end{equation}
where $f(\tau)$ has been defined in Eq.~(\ref{eq:ftau}) and 
\begin{equation}
\label{eq:laplace}
\Delta_{S^2}\Phi
=
\frac{1}{\sin\theta}\partial_{\theta}\left(\sin\theta\partial_{\theta}\Phi\right)
+\frac{1}{\sin^2\theta}\partial^2_{\phi}\Phi\, ,
\end{equation}
is the Laplacian on the round 2-sphere of unit radius.  Using the separation
ansatz
\begin{equation}
\label{eq:separationvariables}
\Phi= e^{-i\omega t} R (\tau) Y^{l}_{m} (\theta,\phi)\, ,
\end{equation}
and
\begin{equation}
\label{eq:delta}
\Delta_{S^2}Y^{l}_{m} (\theta,\phi)=-l(l+1)Y^{l}_{m} (\theta,\phi) \, ,
\end{equation}
we find 
\begin{equation}
\omega^{2} e^{-4 U} f^2 R(\tau)+f^{-2}\partial^2_{\tau}R(\tau) = 
l (l+1)R(\tau)\, ,
\end{equation}
so we can write Eq. (\ref{eq:KG4d}) as
\begin{equation}
\label{eq:KG4dseparated1}
\mathcal{K}_{4}\Phi=l(l+1)\Phi\, ,
\end{equation}
where $\mathcal{K}_{4}$ is the second-order differential operator
\begin{equation}
\label{eq:KG4dH}
\mathcal{K}_{4} \equiv - e^{-4 U} f^{2} \partial^{2}_{t}+f^{-2}\partial^{2}_{\tau}\, .
\end{equation}

In order to exhibit the hidden conformal structure of the given space-time, we
want to find a representation of $SL(2,\mathbb{R})$ in terms of first-order
differential operators (vector fields) in the $t-\tau$ submanifolds, such as
the $SL(2,\mathbb{R})$ quadratic Casimir, constructed from those vector fields
is equal to the second-order differential operator $\mathcal{K}_{4}$. Thus, we
want to find three real vector fields
\begin{equation}
\label{eq:Hs}
L_{m} =  a_{mt}\partial_{t}+a_{m\tau}\partial_{\tau}\, ,
\hspace{1cm}
m=0,\pm 1\, , 
\end{equation}
for some functions $a_{mt}(t,\tau),a_{m\tau}(t,\tau)$, whose Lie brackets are
satisfy $\mathfrak{sl}(2)$ Lie algebra
\begin{eqnarray}
\label{eq:Hsconmutation}
[L_{m},L_{n}]=(m-n)  L_{m+n}\, ,
\hspace{1cm}
m=0,\pm 1\, ,
\end{eqnarray}
and such that 
\begin{eqnarray}
\label{eq:CasimirK}
\mathcal{H}^{2}
\equiv
L^{2}_{0}-\tfrac{1}{2}\left(L_{1} L_{-1}+L_{-1} L_{1}\right) 
= 
\mathcal{K}_{4}\, .
\end{eqnarray}

In order to simplify this problem, following \cite{Bertini:2011ga}, we have to
make some additional assumptions on the functions
$a_{It}(t,\tau),a_{I\tau}(t,\tau)$, Thus, we make the following ansatz
\begin{eqnarray}
\label{eq:Hs2}
L_{1} 
& = & 
l(t)\left[-m(\tau)\partial_{t}+n(\tau)\partial_{\tau}\right]\, ,
\\
& & \nonumber \\
L_{0} 
& = & 
-\frac{c}{r_{0}}\partial_{t}\, ,
\\
& & \nonumber \\
L_{-1} 
& = & 
-l^{-1}(t)\left[ m(\tau)\partial_{t}+n(\tau)\partial_{\tau}\right]\, ,
\end{eqnarray}
where $m$ and $n$ are functions of $\tau$, $l$ is a function of $t$ and $c$ is
a real constant.

Plugging this ansatz into Eq.~(\ref{eq:Hsconmutation}) we obtain two
differential equations
\begin{eqnarray}
\label{eq:diff1}
m^{2}\partial_{t} \log l + n\partial_{\tau} m 
& = & 
\frac{c}{r_{0}}\, ,
\\
& & \nonumber \\
\frac{c}{r_{0}}\partial_{t}\log l = 1\, ,
\end{eqnarray}
and plugging it into Eq.~(\ref{eq:CasimirK}) we obtain three equations
\begin{eqnarray}
\label{eq:diff2}
m 
& = & 
h\partial_{\tau} n\, ,
\\
& & \nonumber \\
m^{2} 
& = &
e^{-4 U} f^{2}+\left(c/r_{0}\right)^{2}\, ,
\\
& & \nonumber \\
n^{2}
& =& f^{-2}\, .
\end{eqnarray}

These equations cannot be solved for arbitrary $U(\tau)$: we can find $l,m,n$
as functions of $f(\tau)$ and the constant $c$
\begin{equation}
\label{eq:sol1}
l(t) = c_{0} e^{r_{0}t/c}\, ,
\hspace{1cm}
 n^{2}(\tau)=f^{-2}\, ,
\hspace{1cm}
m(\tau)=h \cosh{(r_{0}\tau)}\, ,
\end{equation}
for some real constant $c_{0}$, leaving the following equation for the
constant $c$ to be solved:
\begin{eqnarray}
\label{eq:constraint2}
c^{2} = \left(e^{-2U} f^{2}\right)^{2}\, .
\end{eqnarray}

This equation can only be exactly solved, for all values of the radial
coordinate $\tau$ for $e^{U}\sim f$, which does not correspond to any
asymptotically flat black hole. We have to content ourselves with a range of
values of the coordinate $\tau$ in which the above equation can be solved
approximately. The two ranges that we have identified correspond to the two
near-horizon regions (event and Cauchy horizons $\tau\rightarrow -\infty$ or
$\tau\rightarrow -\infty$, respectively) in which 
\begin{equation}
\label{eq:constraint3}
\left(e^{-2U} f^{2}\right)^{2}
\stackrel{\tau \rightarrow \mp \infty}{\sim}
\left(\frac{A_{\pm}}{4\pi}\right)^{2}+\mathcal{O}(e^{\pm r_{0}\tau})
=
c^{2}+\mathcal{O}(e^{\pm r_{0}\tau})\, ,
\end{equation}
according to Eq.~(\ref{eq:areahorizons}).

We conclude that in the geometry of any 4-dimensional, charged, static,
black-hole solution of a theory of the form Eq.~(\ref{eq:generalaction4}),
there are two triplets of vector fields $L^{+}_{m}$ and $L^{-}_{m}$, $m=0,\pm
1$ given by
\begin{eqnarray}
\label{eq:Hs+}
L^{\pm}_{1}  
& = &
-\frac{e^{r_{0}\pi t /S_{\pm}}}{r_{0}}\left(\frac{S_{\pm}}{\pi} \cosh{(r_{0} \tau)}
\partial_{t}+\sinh{(r_{0}\tau)}\partial_{\tau}\right)\\
& & \nonumber \\
L^{\pm}_{0} 
& = &
 -\frac{S_{\pm}}{r_{0}\pi}\partial_{t}\, ,
\\
& & \nonumber \\
L^{\pm}_{-1} 
& = &
-\frac{e^{-r_{0}\pi t /S_{\pm}}}{r_{0}}
\left(\frac{S_{\pm}}{\pi} \cosh{(r_{0}\tau)}\partial_{t} 
-\sinh{(r_{0}\tau)}\partial_{\tau}\right)\, ,
\end{eqnarray}
where $S_{\pm}=\frac{A_{\pm}}{4}$, which generate two $\mathfrak{sl}(2)$
algebras whose quadratic Casimirs
\begin{equation}
\mathcal{H}^{\pm\, 2}
\equiv
(L^{\pm}_{0})^{2}
-\tfrac{1}{2}\left(L^{\pm}_{1} L^{\pm}_{-1}+L^{\pm}_{-1} L^{\pm}_{1}\right)\, ,   
\end{equation}
approximate the massless Klein-Gordon equation in the two near-horizon
regions\footnote{Observe that we only approximate some terms (i.e.~we keep some
  sub-dominating terms): 
\begin{equation}
e^{-4U}f^{2} = f^{-2} (e^{-2U}f^{2})^{2} 
\sim     
f^{-2}
\left[\left(\frac{A_{\pm}}{4\pi}\right)^{2}+\mathcal{O}(e^{\pm
    r_{0}\tau})\right]
\sim f^{-2}
\left(\frac{A_{\pm}}{4\pi}\right)^{2}+\mathcal{O}(e^{\pm
    r_{0}\tau})\, ,
\end{equation}
which is correct to that order. On the other hand, we do not need to restrict
ourselves to any particular range of frequencies.  
}:
\begin{equation}
\mathcal{K}_{4}\Phi 
= \left\{- e^{-4 U} f^{2} \partial^{2}_{t}+f^{-2}\partial^{2}_{\tau} \right\}
\Phi\,\, \stackrel{\tau\rightarrow \mp \infty}{\longrightarrow}\,\,
f^{-2}\left\{-\left(S_{\pm}/\pi \right)^{2}\partial^{2}_{t}+\partial^{2}_{\tau} \right\}
\Phi
=
\mathcal{H}^{\pm\, 2} \Phi\, . 
\end{equation}

We can see from Eq.~(\ref{eq:Hs+}) that the extremal limit $r_{0}\rightarrow
0$ is singular. The reason is that the operations of taking the near-horizon
limit and of taking the extremal limit $r_{0}\rightarrow 0$ do not commute.

The $\mathfrak{sl}(2)$ algebra that we have just found can be immediately
extended to a complete Witt algebra (or a Virasoro algebra with vanishing
central charge) with the commutation relations (\ref{eq:Hsconmutation}) for
all $m\in \mathbb{Z}$. The generators of the Witt algebra are given by
\begin{eqnarray}
\label{eq:Lm}
L^{\pm}_{m}  
& = &
-\frac{e^{m r_{0}\pi t /S_{\pm}}}{r_{0}}\left(\frac{S_{\pm}}{\pi} \cosh{(m r_{0} \tau)}
\partial_{t}+\sinh{(m r_{0}\tau)}\partial_{\tau}\right) \, .
\end{eqnarray}


\section{Hidden conformal symmetry in $d$ dimensions.}

We are now ready to generalize the results of the previous section to to
arbitrary $d\geq 4$ dimensions, using the general metric
Eq.~(\ref{eq:generalbhmetricd}). In this background, the massless Klein-Gordon
equation can be written as
\begin{equation}
\label{eq:KGd}
e^{-\frac{2(d-2)}{(d-3)}U }g^{\frac{2}{d-3}}\partial^{2}_{t}\Phi
-(d-3)^{2}g^{-2}\partial^{2}_{\rho}\Phi
-\Delta_{S^{d-2}}\Phi=0\, ,
\end{equation}
where $g(\rho)$ is defined in Eq.~(\ref{eq:grho}) and $\Delta_{S^{d-2}}$ is
the Laplacian in the round $(d-2)$ -sphere of unit radius. Using the
separation ansatz
\begin{equation}
\Phi= e^{-i\omega t} R (\rho) Y^{l}_{\mu} (\vec{\theta})\, ,
\end{equation}
where $Y^{l}_{\mu}(\vec{\theta})$ are the spherical harmonics on
$S^{d-2}$, Eq.~(\ref{eq:KGd}) takes the form
\begin{equation}
\label{eq:KGd2}
\frac{e^{-\frac{2(d-2)}{(d-3)}U }}{(d-3)^{2}}g^{\frac{2}{d-3}}\omega^{2}R(\rho)
+g^{-2}\partial^{2}_{\rho}R(\rho)=\frac{l(l+d-3)}{(d-3)^{2}}R(\rho)\, ,
\end{equation}
so the Klein-Gordon equation takes the form 
\begin{equation}
\mathcal{K}_d\Phi=\frac{l(l+d-3)}{(d-3)^{2}}\Phi\, .
\end{equation}
where we have defined the reduced Klein-Gordon operator $\mathcal{K}_{d}$
\begin{equation}
\label{eq:Kd2}
\mathcal{K}_d=-\frac{e^{-\frac{2(d-2)}{(d-3)}U }}{(d-3)^{2}}
g^{\frac{2}{d-3}}\partial^{2}_{t}
+g^{-2}\partial^{2}_{\rho}\, .
\end{equation}

As in the four-dimensional case, we want to find two triplets of vector fields
generating the $\mathfrak{sl}(2)$ Lie algebra and whose quadratic Casimir
approximates the $d$-dimensional reduced Klein-Gordon operator
$\mathcal{K}_{d}$ in some region of the black-hole spacetime. It is not too
hard to show that the two triplets
\begin{eqnarray}
\label{eq:Ld}
L^{\pm}_{1}  
& = &
-\frac{e^{(d-3)C_{(d-2)}\mathcal{B} t /A_{\pm}}}{\mathcal{B}}
\left(\frac{A_{\pm}}{(d-3)C_{(d-2)}} \cosh{(\mathcal{B} \tau)}
\partial_{t}
+\sinh{(\mathcal{B}\tau)}\partial_{\tau}\right)\\
& & \nonumber \\
L^{\pm}_{0} 
& = &
 -\frac{A_{\pm}}{(d-3)C_{(d-2)}\mathcal{B}}\partial_{t}\, ,
\\
& & \nonumber \\
L^{\pm}_{-1} 
& = &
-\frac{e^{-(d-3)C_{(d-2)}\mathcal{B} t /A_{\pm}}}{\mathcal{B}}
\left(\frac{A_{\pm}}{(d-3)C_{(d-2)}} \cosh{(\mathcal{B}\tau)}\partial_{t} 
-\sinh{(\mathcal{B}\tau)}\partial_{\tau}\right)\, ,
\end{eqnarray}
where Eqs. (\ref{eq:arearho}) and (\ref{eq:areahorizond}) have been used in
order to take the near horizon $\rho\rightarrow\pm\infty$ limit.

Extending these two $\mathfrak{sl}(2)$ algebras to two full Witt algebras is
straightforward:
\begin{eqnarray}
\label{eq:Lmd}
L^{\pm}_{m}  
= 
-\frac{e^{m(d-3)C_{(d-2)}\mathcal{B} t /A_{\pm}}}{\mathcal{B}}
\left(\frac{A_{\pm}}{(d-3)C_{(d-2)}} \cosh{(m\mathcal{B} \tau)}
\partial_{t}
+\sinh{(m\mathcal{B}\tau)}\partial_{\tau}\right) \, .
\end{eqnarray}


\section{Discussion}
\label{sec-discussion}

In this paper we have constructed two Witt algebras which have a well-defined
action in the space of solutions to the wave equation in the background of the
exterior and interior near-horizon limits of a generic, charged, static black
hole. The two $\mathfrak{sl}(2)$ subalgebras are symmetries of these wave
equations, since the wave operators can be seen as their Casimirs, but they
are not symmetries of the background metrics which, being essentially the
products of Rindler spacetime (locally Minkowski) and spheres, have Abelian
(in the time-radial part) isometry algebras.

This result generalizes those obtained in
\cite{Bertini:2011ga,Wang:2010qv,Lowe:2011aa,Camblong:2004ye}, and present an
opportunity to put to test some conjectures and common lore of this field. To
start with, is there a CFT associated to the Witt algebras and can one compute
the central charge of the Virasoro algebra? A most naive computation does not
seem to give meaningful results. This, of course, does not preclude the
possibility that a more rigorous calculation, preceded of careful definitions
of the boundary conditions of the fields at the relevant boundaries (which
have to be identified first) may give a meaningful answer.

Meanwhile, it is amusing to speculate on the possible consequences of the
existence of such a CFT withe left and right sectors whose entropies $S_{\rm
  R},S_{\rm L}$ and temperatures would be related to the temperatures and
entropies of the outer and inner horizons ($T_{+},T_{-}$ and $S_{+},S_{-}$,
respectively) by

\begin{eqnarray}
S_{\pm} & = & S_{\rm R} \pm S_{\rm L}\, , \\
& & \nonumber \\
\frac{1}{T_{\pm}} & = & 
\tfrac{1}{2}
\left(
\frac{1}{T_{\rm R}} \pm \frac{1}{T_{\rm L}}
\right)\, , 
\end{eqnarray}

\noindent
and obeying the fundamental relation 

\begin{equation}
S_{+} = \frac{\pi^{2}}{12}(c_{\rm R}T_{R} +c_{\rm L}T_{\rm L})\, ,  
\end{equation}

\noindent
where $c_{\rm L,R}$ are the central charges of the left and right sectors,
which will be assumed to be equal $c_{\rm R}= c_{\rm L} = c$.

The temperatures and entropies of the outer and inner horizons are related to
the non-extremality parameter $t_{0}$ by

\begin{equation}
2S_{\pm}T_{\pm} = r_{0}\, ,  
\end{equation}

\noindent
which implies for the temperatures of the left and right sectors

\begin{equation}
4S_{\rm L,R}T_{\rm L,R} = r_{0}\, .  
\end{equation}

In the extremal limit 

\begin{equation}
S_{\rm L} \rightarrow 0\, ,
\hspace{.5cm}
T_{\rm R} \rightarrow 0\, ,
\hspace{.5cm}
T_{\pm} \rightarrow 0\, ,
\hspace{.5cm}
S_{\pm} \rightarrow S_{\rm R}\, ,
\end{equation}

\noindent
and both $S_{\rm R}$ and $T_{\rm L}$ remain finite and are convenient
quantities to work with. In particular, we can express the central charge that
the CFT should have in order to reproduce the Bekenstein-Hawking entropy 
consistently with this picture, in terms of these two parameters:

\begin{equation}
c = \frac{12}{\pi^{2}} \frac{S_{\rm R}}{T_{\rm L}}\, .  
\end{equation}


\section*{Acknowledgments}

The authors would like to thank Patrick Meessen for very useful conversations.
This work has been supported in part by the Spanish Ministry of Science and
Education grant FPA2009-07692, the Comunidad de Madrid grant HEPHACOS
S2009ESP-1473 and the Spanish Consolider-Ingenio 2010 program CPAN
CSD2007-00042. The work of CSS has been supported by a CSIC JAE-predoc grant
JAEPre 2010 00613. TO wishes to thank M.M.~Fern\'andez for her permanent
support.



\begin{thebibliography}{99}

\raggedright

\bibitem{Bertini:2011ga}
S.~Bertini, S.~L.~Cacciatori and D.~Klemm,
\arxiv{1106.0999} [hep-th].

\bibitem{Strominger:1996sh}
A.~Strominger, C.~Vafa,
Phys.\ Lett.\  {\bf B379} (1996)  99-104
[\hepth{9601029}]. 

\bibitem{Compere:2012jk}
G.~Comp\`ere,
\arxiv{1203.3561} [hep-th].

\bibitem{Strominger:1997eq}
A.~Strominger,
JHEP\ {\bf 9802} (1998) 009
[\hepth{9712251}].

\bibitem{Banados:1992wn}
M.~Ba\~nados, C.~Teitelboim and J.~Zanelli,
Phys.\ Rev.\ Lett.\ \ {\bf 69} (1992) 1849
[\hepth{9204099}].

\bibitem{Brown:1986nw}
J.~D.~Brown and M.~Henneaux,
Commun.\ Math.\ Phys.\ \ {\bf 104} (1986) 207.

\bibitem{Carlip:1998qw}
S.~Carlip,
Class.\ Quant.\ Grav.\ \ {\bf 15} (1998) 3609
[\hepth{9806026}].

\bibitem{Guica:2008mu}
M.~Guica, T.~Hartman, W.~Song and A.~Strominger,
Phys.\ Rev.\  D {\bf 80} (2009) 124008
[\arxiv{0809.4266} [hep-th]].

\bibitem{Castro:2010fd}
A.~Castro, A.~Maloney and A.~Strominger,
Phys.\ Rev.\  D {\bf 82} (2010) 024008
[\arxiv{1004.0996} [hep-th]].

\bibitem{Wang:2010qv}
Y.~Q.~Wang and Y.~X.~Liu,
JHEP {\bf 1008} (2010) 087
[\arxiv{1004.4661} [hep-th]].

\bibitem{Camblong:2004ye}
H.~E.~Camblong and C.~R.~Ordonez,
Phys.\ Rev.\ D {\bf 71} (2005) 104029
[\hepth{0411008}].

\bibitem{Ferrara:1997tw}
S.~Ferrara, G.~W.~Gibbons, R.~Kallosh,
Nucl.\ Phys.\  {\bf B500} (1997)  75-93
[\hepth{9702103}].

\bibitem{Meessen:2011bd}
P.~Meessen and T.~Ort\'{\i}n,
Phys.\ Lett.\ {\bf B707} (2012) 178
[\arxiv{1107.5454}].

\bibitem{Govindarajan:2000ag}
T.~R.~Govindarajan, V.~Suneeta and S.~Vaidya,
Nucl.\ Phys.\ B\ {\bf 583} (2000) 291
[\hepth{0002036}].

\bibitem{Birmingham:2001qa}
D.~Birmingham, K.~S.~Gupta and S.~Sen,
Phys.\ Lett.\ B\ {\bf 505} (2001) 191
[\hepth{0102051} [hep-th]].

\bibitem{Gupta:2001bg}
K.~S.~Gupta and S.~Sen,
Phys.\ Lett.\ B\ {\bf 526} (2002) 121
[\hepth{0112041}].

\bibitem{Franzin:2011wi}
E.~Franzin and I.~Smolic,
JHEP\ {\bf 1109} (2011) 081
[\arxiv{1107.2756} [hep-th]].

\bibitem{Galli:2011fq}
P.~Galli, T.~Ort\'{\i}n, J.~Perz, C.~S.~Shahbazi,
JHEP {\bf 1107} (2011) 041
[\arxiv{1105.3311}].

\bibitem{Meessen:2012su}
P.~Meessen, T.~Ort\'{\i}n, J.~Perz and C.~S.~Shahbazi,
\arxiv{1204.0507} [hep-th].

\bibitem{Lowe:2011aa}
D.~A.~Lowe and A.~Skanata,
\arxiv{1112.1431} [hep-th].


\end{thebibliography}
\end{document}